\documentclass[runningheads]{llncs}
\usepackage{graphicx}
\usepackage{hyperref}
\begin{document}
\title{Dealing with Data for RE: Mitigating Challenges while using NLP and Generative AI}
\titlerunning{Dealing with Data for RE}
%
\author{
Smita Ghaisas \and Anmol Singhal} 
\authorrunning{Ghaisas and Singhal}
%
\institute{TCS Research, India}
\maketitle              
\begin{abstract}
Across the dynamic business landscape today, enterprises face an ever-increasing range of challenges. These include the constantly evolving regulatory environment, the growing demand for personalization within software applications, and the heightened emphasis on governance. In response to these multifaceted demands, large enterprises have been adopting automation that spans from the optimization of core business processes to the enhancement of customer experiences. Indeed, Artificial Intelligence (AI) has emerged as a pivotal element of modern software systems. In this context, data plays an indispensable role. AI-centric software systems based on supervised learning and operating at an industrial scale require large volumes of training data to perform effectively. Moreover, the incorporation of generative AI has led to a growing demand for adequate evaluation benchmarks. Our experience in this field has revealed that the requirement for large datasets for training and evaluation introduces a host of intricate challenges. This book chapter explores the evolving landscape of Software Engineering (SE) in general, and Requirements Engineering (RE) in particular, in this era marked by AI integration. We discuss challenges that arise while integrating Natural Language Processing (NLP) and generative AI into enterprise-critical software systems. The chapter provides practical insights, solutions, and examples to equip readers with the knowledge and tools necessary for effectively building solutions with NLP at their cores. We also reflect on how these text data-centric tasks sit together with the traditional RE process. With this effort, we hope to engage students, faculty, and industry researchers in a discussion that could lead to the identification of new and emerging text data-centric tasks relevant to RE. We also highlight new RE tasks that may be necessary for handling the increasingly important text data-centricity involved in developing software systems.
\end{abstract}
\section{Introduction}

In recent times, we have witnessed a historic transformation in Software Engineering (SE) practices due to the growing prominence of AI as a key constituent of modern software systems. Most customer-centric enterprises in the technology sector are now integrating AI-powered solutions into their product and service offerings \cite{ref1}. Moreover, Natural Language Processing (NLP) is increasingly being used to automate tasks within the Software Development Lifecycle (SDLC) \cite{ref2}, particularly, those tasks relevant to Requirements Engineering (RE). As requirements are mainly captured in Natural Language (NL), the prominent use of NLP is easy to appreciate. Additionally, as industry researchers, we observe that enterprises are seeking AI-based solutions to process the enormous amount of text data present in business-critical documents besides Software Requirements Specifications (SRS). These include contracts, regulations, and privacy policies.

Furthermore, as we transition into the era of Artificial General Intelligence (AGI), Large Language Models (LLMs) are seen demonstrating a remarkable prowess in interpreting Natural Language (NL). These models can not only learn to comprehend NL but can also acquire the capability to reason over content. As a result, forward-thinking SE enterprises, in collaboration with businesses seeking automation, are integrating generative AI models into business processes and product development pipelines \cite{ref3}. This evolution across the software landscape has made text data an integral component of SE. 

In light of this transforming landscape, researchers have been focusing on modeling data requirements for AI-centric software solutions \cite{Vogelsang2019RequirementsEF,Altarturi2017ARE,sambavisan}. While some of the findings of the works apply to text data-centric projects, research dedicated to identifying specific data requirements for text data-centric software remains an open area. This emphasis on text data is essential to success in building robust text data-centric software systems. 

In this chapter, we focus on text data-centric challenges likely to be encountered while integrating NLP into SE. We refer to the resulting systems as Software Systems with NLP at the Core (SSNLPCore). These challenges are particularly pronounced in the development of industry-oriented software solutions, which demand scalability, reliability, and robustness against variations in data distribution that may change over time and across geographies. The RE community (academia and industry alike), which is now operating at the intersection of SE, NLP, and Generative AI, needs to focus on the following Research Questions (RQs) and collaborate to address them: (1) What are the data-centric challenges encountered in SSNLPCores? (2) How do the text data-centric challenges integrate with the traditional RE process we are familiar with so far? 

We note that we do not have precise answers at this point. Nevertheless, we have some insights to share with the RE community, based on a review of existing literature, our earlier empirical work \cite{cain} on identifying data-centric challenges in building SSNLPCores, and our continued focus on attention to data-centricity. 

The chapter aims to provide insights into the above-mentioned RQs. To answer RQ1, we first describe the stages in an SSNLPCore project lifecycle where data-centric challenges typically occur. We then elaborate on the solution/s that are being adopted by practitioners to mitigate the commonly encountered data-centric challenges in each stage and illustrate the application of these solutions on publicly available datasets. Subsequently, we answer RQ2 by: (1) Detailing the implications of the data-centricity of SSNLPCores on the traditional RE process, and (2) Highlighting the relevance of data-centricity while automating tasks relevant to the RE process. We aim to provide practical insights into the complexities and challenges involved in integrating NLP and Generative AI within software solutions. We hope to equip you with the knowledge and strategies necessary for informed decision-making and for effectively addressing data-centric challenges. Finally, our objective is to initiate discussions in the RE community about the ‘entry points’ for the data-centric tasks into the traditional RE process for developing SSNLPCores. 

The rest of the chapter is organized as follows: Section \ref{s2} provides an overview of the data-centric challenges encountered in the different stages of building SSNLPCores. Section \ref{s3} elaborates on the publicly available datasets used for examples in this chapter.  This section also details the mitigation strategies for data-centric challenges that can occur while building an SSNLPCore project. In Section \ref{s4}, we discuss the impact of the data-centricity of SSNLPCores on the traditional RE process and identify some possible `entry points’. Section \ref{s5} concludes the chapter.  
\vspace{-3mm}
\section{Data-centric Stages in an SSNLPCore Lifecycle}
\label{s2}
The majority of cutting-edge approaches in NLP leverage Deep Learning (DL)-based models, and they need a large amount of training data \cite{cain}. Consequently, teams involved in building SSNLPCores are compelled to allocate a significant portion of their time to the meticulous gathering, processing, and analysis of large datasets. Additionally, when it comes to deploying SSNLPCores in real-world scenarios, it becomes important to concentrate on the generalizability and trustworthiness of the solutions. Efforts directed at ensuring that SSNLPCores meet these requirements need access to high-quality datasets. However, collecting datasets with the desired attributes can entail a unique set of challenges. 

Data-centric challenges occur during data collection, annotation, processing, and validation \cite{cain,ref8}. In this section, we detail each stage by highlighting the significant role each stage plays in the overall development lifecycle of an SSNLPCore and the potential data-centric challenges that can arise while executing tasks in each stage. The identified stages and challenges are based on our earlier empirical investigation \cite{cain}.

As an example, we also illustrate the data-centric challenges that may occur while building a solution for education feedback analysis. In the education domain, student feedback data is necessary to uncover the merits and demerits of existing services provided to students. An NLP-centric system can play a vital role in analyzing student feedback in text format. This example is borrowed from the literature review by Shaik et al. \cite{shaik}. 

\subsection{Data Collection}

Data collection constitutes the first stage in the development of SSNLPCores after the identification of the problem to be solved. The key challenges that can occur during data collection are:

\begin{itemize}
    \item \textit{Limited availability of data sources}: A scarcity of sources from which data can be extracted can limit the size of datasets, adversely affecting the model's ability to learn complex patterns.
    \item \textit{Class imbalance}: Class imbalance can lead to biased learning towards a particular class, based on the distribution of samples in the dataset. 
    \item \textit{Complexity of the business domain in which the system is to be deployed}: If the business domain in which the system is to be deployed is highly complex, such as law or medicine, it necessitates the collection of data that is sufficiently detailed and domain-specific. However, access to domain-specific datasets for training DL-based models is typically constrained. 
    \item \textit{Societal biases}: Societal biases in data can alter the model predictions against a certain section of the population, making it a matter of concern for systems deployed in the real world. Biased models can inadvertently perpetuate and amplify these biases in society.
    \item \textit{Inaccurate data}: Inaccurate training data containing fake examples can hinder the model’s reliability and lead to misinformation.
    \item \textit{Representativeness of the data collected}: If the collected data is not representative of the target domain, it can hamper the ability of SSNLPCores to generalize well after deployment. 
\end{itemize}

In the context of an education feedback analysis system, Eggert et al. \cite{eggert2021artificial} suggest the importance of collecting vast amounts of data related to each student's prior knowledge, emotional state, or economic background. However, collecting such data is prone to issues such as bias, lack of representativeness, and misinformation. Furthermore, in the education domain, it is difficult to collect labeled data as it requires manual annotation from domain experts. In addition, the classification results are often biased and poor due to imbalance and data distribution discrepancies. 
\vspace{-3mm}
\subsection{Data Annotation}
Documents and text data obtained from the web are predominantly unlabeled. Given that the majority of the architectures employed for building traditional SSNLPCores rely on supervised learning, the annotation of unlabeled data is an indispensable requirement for these projects. The annotators engaged in the meticulous annotation exercise can either be Subject Matter Experts (SMEs) or researchers/engineers who are adeptly trained or mentored by SMEs. The key challenges that can emerge in this phase include:

\begin{itemize}
    \item \textit{Subjectivity}: It refers to the potential for varying interpretations and labels assigned by different annotators, affecting the consistency of the dataset.
    \item \textit{Lack of credibility}: Involving annotators with insufficient domain expertise can negatively affect the quality of labeled data and downgrade the intended results of SSNLPCores. 
    \item \textit{Cognitive burden associated with manually annotating datasets}: The excessive cognitive load associated with annotating extensive amounts of data can lead to poor annotation quality and increase the time, effort, and money required for this stage. 
\end{itemize}

Data annotation for building an education feedback analysis system requires manual effort and intervention from domain experts. The annotation can involve document-level categorization, entity extraction, and sentiment annotation of the collected datasets for analysis \cite{shaik}. Therefore, annotating domain-specific datasets for each task in this domain is often prone to errors and subjectivity. 
\vspace{-3mm}
\subsection{Data Processing}

The analysis and processing of the collected data is an important stage that precedes sending the data as input to the AI model. This stage principally encompasses actions necessary for cleaning the data and converting it into a format that is according to the model's expectations. The primary challenges that can emerge during this stage are:
\vspace{-2mm}
\begin{itemize}
    \item \textit{Unstructured nature of text data sources}: The unstructured nature of sources from which text data is extracted demands intricate processing algorithms capable of extracting meaningful information.
    \item \textit{Specialized language and complexity of the text}: They necessitate an understanding of domain-specific jargon.
    \item \textit{Lexical and syntactic errors}: These issues entail handling variations in terminology, grammar, and structure while ensuring that the meaning of the content is retained.
    \item \textit{Operational restrictions, such as confidentiality and compliance-related concerns}: They necessitate the judicious handling of data to meet legal and ethical standards. 
\end{itemize}

Shaik et al. \cite{shaik} elaborate on the domain-specific challenges for building an education feedback analysis tool. These challenges include the presence of jargon, ambiguity, and sarcasm within the text. Furthermore, information such as the economic background of students can be confidential and hard to obtain.
\vspace{-3mm}
\subsection{Data Validation}

This stage involves the validation of the output or the data generated as a consequence of employing SSNLPCores. The principal challenges associated with this stage include:
\vspace{-2mm}
\begin{itemize}
    \item \textit{Lack of evaluation benchmarks to determine generalizability}: The absence of established evaluation benchmarks can undermine the assessment of a model's performance concerning industry standards.
    \item \textit{Inconsistent criteria to determine trustworthiness}: Inconsistent criteria for the trustworthiness of the system make it difficult to establish a consensus on the reliability and credibility of the system. 
    \item \textit{Quantification of a solution's scalability}: Measuring scalability is essential to ascertain the system's capacity to efficiently handle an escalating volume of data or complexity without a degradation in performance.
\end{itemize}

Shaik et al. \cite{shaik} observe the significance of text analytics models for evaluating text data in the education domain. These analysis techniques can include text categorization, entity extraction, sentiment analysis, and document summarization. However, utilizing these analytics models is conditioned on the underlying effort required in data annotation and processing. Furthermore, the biased nature of data hampers the trustworthiness and generalizability of the solution.

Going forward, we describe the mitigation strategies to deal with data-centric challenges in each stage highlighted above. We also present some examples for readers to try on publicly-available datasets. 
\vspace{-3mm}
\section{Mitigating Data-Centric Challenges}
\label{s3}
\vspace{-1mm}
We now elaborate on some mitigation strategies and guidelines used by practitioners to deal with data-centric challenges in different stages of building an SSNLPCore. First, we describe the publicly available datasets used to carry out the examples in the section. Subsequently, we categorize the mitigation strategies into two groups: (1) data collection and annotation, and (2) data processing and validation. As a practical use case, we also discuss an end-to-end data management flow for the education feedback analysis system mentioned above.  
\vspace{-2mm}
\subsection{Datasets used in Exercises and/or Examples}

The datasets used to illustrate the application of state-of-the-art NLP techniques for addressing the challenges are publicly available. Their details are as follows:

\begin{itemize}
    \item \textbf{Expanded PROMISE dataset}: The PROMISE repository is the most commonly used database \cite{Sayyad-Shirabad+Menzies:2005,ref10} for requirements classification into functional and non-functional. This repository is unbalanced and has only 625 labeled requirements written in natural language. To increase the size of the dataset, an expanded version of the PROMISE dataset was released by Lima et al. \cite{ref11}. It contains a fine-grained labeling of Non-Functional Requirements (NFRs). It divides the given 969 NFRs into 12 different classes. 
    \item \textbf{PURE dataset} \cite{ferrari_alessio_2018_1414117}: The dataset was released to contribute to research in natural language requirement processing. It consists of 79 publicly available natural language requirements documents collected from the Web. The dataset includes 34,268 sentences and can be used for NLP tasks that are typical in RE, such as model synthesis, abstraction identification, and document structure assessment. It can be further annotated to perform tasks such as requirements classification and ambiguity detection. 
    \item \textbf{Unfair-TOS Dataset} \cite{Lippi_2019}: The dataset was released to automate the detection of potentially unfair clauses in Terms of Service (ToS) agreements. It contains clauses extracted from 50 ToS agreements of online platforms and labeled as fair or potentially unfair. The contracts were selected among those offered by some of the major players in terms of number of users, global relevance, and time of establishment of the service. The final corpus consists of 12,011 sentences, out of which 1,032 sentences were labeled potentially unfair. The potentially unfair sentences were further labeled into 8 sub-categories depending on the reason behind unfairness. The sub-categories included Arbitration, Unilateral change, Content removal, Jurisdiction, Choice of law, Limitation of liability, Unilateral termination, and Contract by using. The annotation procedure, definitions, and examples of each type of unfair clause are provided in the paper \cite{Lippi_2019}. 
    \item \textbf{LEXDEMOD Dataset \cite{sancheti2022agentspecific}}: The dataset was released to contribute to research on deontic modality detection in commercial contracts written in the English language. The contractual sentences in the dataset were sourced from the LEDGAR corpus \cite{tuggener-etal-2020-ledgar}. The sentences are annotated with deontic modality (obligation, entitlement, prohibition, permission, no obligation, no entitlement) expressed with respect to a contracting party or agent along with the modal triggers. The dataset contains 8,230 span annotations for 7,092 sentences from 23 lease contracts. 
\end{itemize} 

We note here that we limit the scope of our examples to publicly available data and are unable to provide industry data for the examples in this chapter, owing to the confidential nature of industry data. However, the findings of the chapter are based on validating them on large industry datasets. We have cited the necessary references for each finding and provided insights from our empirical study conducted with 18 professionals with extensive industry experience \cite{cain}. The code for all the examples can be found on GitHub \footnote{\href{https://github.com/anmolsinghal98/NLP4RE-Dealing-with-Data}{GitHub Repository}}. 

\subsection{Data Collection and Annotation Stages}
\label{s3.2}
\subsubsection{3.2.1 Ensuring Data Availability} 

While there are curated datasets for conducting research across domains, access to labeled, high-quality data is limited when it comes to implementing SSNLPCores in the real world. The unavailability of data for training DL models constitutes a pressing obstacle that impedes the successful implementation of SSNLPCores. Over time, numerous practitioners have suggested diverse techniques aimed at alleviating this challenge. In the following section, we discuss some of the most frequently employed strategies.

 \paragraph{A) Transfer Learning: }

Over the years, a pivotal advancement in the realm of DL and NLP research has been the emergence of transfer learning \cite{ruder-etal-2019-transfer}. As DL-based models evolved in size and complexity, collecting data of sufficient magnitude to conduct specialized tasks grew progressively challenging. The research community addressed this concern by conceptualizing the mechanism of pre-training and fine-tuning, commonly referred to as transfer learning.

To explicate, transfer learning can be broken down into two sequential steps: In the first phase, known as pre-training, the fundamental DL-based neural network undergoes training on a vast corpus of general-purpose text sources, such as news articles and social media content, utilizing a language modeling objective. Subsequently, during the fine-tuning phase, this pre-trained model undergoes further training on a more modest, task-specific corpus. This mechanism efficiently minimizes the necessity for a large, specialized dataset for each task, a resource that is often arduous to collect, compared to the ubiquity of general-purpose text. Additionally, it also enhances the model's competence to effectively handle downstream tasks.

With the advent of pre-trained language models like BERT \cite{devlin-etal-2019-bert} and GPT-2 \cite{Radford2019LanguageMA}, transfer learning has gained widespread acceptance among practitioners spanning various domains, including RE. Over the last few years, RE researchers have conducted extensive experimentation with transfer learning, showcasing its versatility across a diverse array of tasks. In the exercise given below, we elucidate the application of transfer learning for a multi-class classification task, thereby providing practical insights into this technique.\\

\textit{Example 1}: We illustrate the effectiveness of transfer learning on the expanded PROMISE dataset. The task is to automate the classification of a given requirement for an SSNLPCore system into functional or non-functional (binary classification). We conducted the following steps:
\begin{enumerate}
    \item We divided the dataset into an 80:20 train-test split using random sampling.
    \item We used the training data to fine-tune a pre-trained model. In this case, we used BERT (the pre-trained version of BERT named bert-base-cased is available on the Huggingface library). 
    \item We alternately trained a vanilla Transformer model using Pytorch. 
    \item We evaluated both models trained in steps 2 and 3 on the test set.
\end{enumerate}

\textit{Observations}: Fine-tuned BERT outperformed the Transformer architecture trained from scratch by a significant margin of 35\%. BERT performed the binary classification with an accuracy of approximately 90\% and an F1 score of 0.90 on the test set. Please note here that the exact margin by which fine-tuned BERT outperforms the vanilla Transformer model depends upon the hyperparameters used. The hyperparameters we used in this example are listed in Table \ref{tab1}. Therefore, the results may vary if the hyperparameters are changed.
\vspace{-4mm}
\begin{table}
 \renewcommand{\arraystretch}{1.2}
\centering
\scriptsize
\caption{Hyperparameters Used}
\label{tab1}
\begin{tabular}{ p{4cm} p{2cm} c}
\hline
Hyperparameter &  Value \\
\hline
No. of Epochs & 20 \\
Learning Rate	& 2e-5 \\
Max Input Length & 160 \\
Batch Size	& 16 \\
Dropout	& 0.3 \\
\hline
\end{tabular}
\vspace{-3mm}
\end{table}
\vspace{-3mm}

\paragraph{B) Prompting: }
A significant milestone in the evolution of NLP research is the advent of the pre-training, prompting, and predicting paradigm, more succinctly referred to as `prompting' \cite{liu2021pretrain}.

LLMs, trained on an immense volume of data, exhibit exceptional capabilities in reasoning and inference. Predominantly, these LLMs function as decoder-only models, generating tokens based on specified prompts or instructions. These prompts provide the contextual foundation for the LLM, influencing the generation of subsequent tokens. Furthermore, the efficiency of an LLM can be augmented by supplementing prompts with in-context examples, enabling it to comprehend the intended task and deliver outputs more effectively. When a prompt includes examples for a specified task, this configuration is termed `few-shot prompting.' Conversely, if no examples are furnished, it is characterized as `zero-shot prompting' \cite{brown2020language}.

The primary advantage of prompting lies in its ability to reduce the need for training or fine-tuning a language model. There is no requirement to compile large datasets for individual tasks. Instead, one can instruct the LLM to execute the task by providing the requisite prompts, and, if needed, enhance the generation quality by adding a few examples. However, it should be noted that prompting has not entirely obviated the need for training or fine-tuning language models, as LLMs typically struggle to generalize well on unseen and specialized tasks \cite{ref20}. Despite this limitation, prompting has initiated new research avenues, with practitioners increasingly focusing on developing innovative prompting mechanisms to tackle complex tasks. The number of tokens that can be integrated into a prompt depends on the context window of the LLM utilized. In the forthcoming exercise, we will illustrate the mechanism of prompting. \\

\textit{Example 2}: The goal is to repeat the requirement classification task given in example 1 using prompting. We used the expanded PROMISE dataset and conducted the following steps:

\begin{enumerate}
    \item We used the OpenAI API \footnote{\href{https://openai.com/blog/openai-api}{OpenAI API}} and the Langchain library  \footnote{\href{https://docs.langchain.com/docs/}{Langchain}} to perform the task.
    \item We prompted ChatGPT (gpt-3.5-turbo model) to predict the classification labels for the requirement provided. We experimented with different prompts in the few-shot and zero-shot settings. 
    \item We repeated step 2 for all the requirements present in the dataset and obtained the model’s overall accuracy and F1 score. 
\end{enumerate} 

\textit{Observations}: The prompts and the results obtained are provided in Table \ref{tab2}. ChatGPT can exhibit substantial accuracy on the classification task, given the right prompt is provided. Note that you need to create an OpenAI account in order to access the API. Alternately, open-source LLMs available on Huggingface \cite{wolf2020huggingfaces}, such as LLAMA \cite{touvron2023llama} and Vicuna \cite{vicuna2023}, can be used for these experiments. 

\begin{table}[t]
 \renewcommand{\arraystretch}{1.2}
\centering
\scriptsize
\caption{Prompts and Results for Example 2}
\label{tab2}
\begin{tabular}{ p{8cm} p{1.5cm} c c}
\hline
Prompt &  Setting & Accuracy & F1-Score\\
\hline
Classify the following requirement as “Functional” or “Non-Functional”. \\
Examples: $<$\textit{list}$>$ \\
Requirement: \textit{$<$input$>$}	& Few-Shot (n=2)	& 76.3\% & 0.76\\
\hline
You are a requirements analyst. Classify the given software requirement as “Functional” or “Non-Functional”. Please read each requirement carefully and determine its classification based on the definitions provided below. \\
Functional: $<$\textit{definition}$>$ \\
Non-Functional: $<$\textit{definition}$>$ \\
Requirement: $<$\textit{input}$>$	& Zero-Shot	& 82.9\% & 0.83\\
\hline
You are a requirements analyst. Classify the given software requirement as “Functional” or “Non-Functional”. Please read each requirement carefully and determine its classification based on the definitions provided below. \\
Functional: $<$\textit{definition}$>$ \\
Non-Functional: $<$\textit{definition}$>$ \\
Examples: $<$\textit{list}$>$ \\
Requirement: $<$\textit{input}$>$	& Few-Shot	(n=2) & 87.1\% & 0.87\\
\hline
\end{tabular}
\vspace{-4mm}
\end{table}

\paragraph{C) Self-Training: }

While unlabeled data is often available in abundance, securing access to a labeled dataset frequently poses a challenge. Researchers have introduced a plethora of semi-supervised and unsupervised machine learning algorithms to capitalize on the availability of extensive unlabeled datasets. In this section, we highlight one of the most straightforward yet effective semi-supervised paradigms, known as self-training \cite{amini2023selftraining,singhal-etal-2023-towards}.

The self-training principle operates by iteratively training a classifier, assigning pseudo-labels to a subset of the unlabeled training data for which prediction confidence exceeds a particular threshold. These pseudo-labeled samples are then utilized to enrich the labeled training data, upon which a newly trained classifier is based, thus incorporating both labeled and pseudo-labeled data. Typically, the self-training framework encompasses three main steps: 1) Training a language model, referred to as the Teacher, for the intended task using solely the labeled training data. 2) Obtaining predictions on unlabeled data and filtering them based on a confidence threshold criterion. Unlabeled sentences meeting this criterion are incorporated into the training set, with their predictions treated as pseudo-labels. 3) Using the augmented dataset, comprised of both labeled and pseudo-labeled sentences, training a new model called the Student.

\begin{figure*}[!h]
\centering
\includegraphics[scale=0.80]{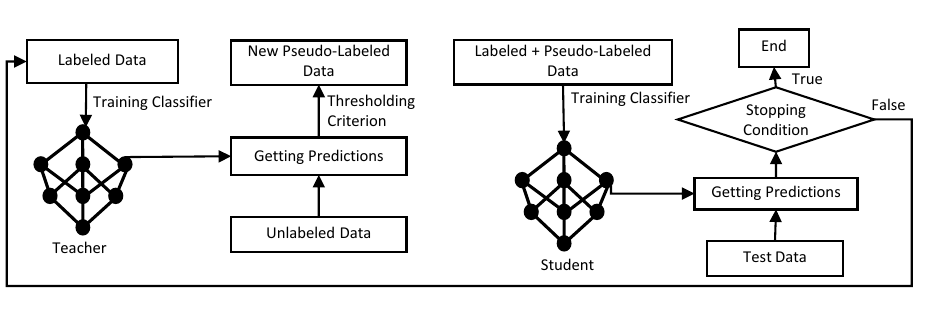}
\vspace{-5mm}
\caption{Self-Training Framework}
\label{fig1}
\vspace{-8mm}
\end{figure*}

These steps are iteratively repeated until no unlabeled sentences remain. In each subsequent cycle, pseudo-labels are assigned to the remaining unlabeled data to train a fresh model. The self-training framework is depicted in Figure \ref{fig1}. \\

\textit{Example 3}: We illustrate self-training by using a combined dataset of labeled and unlabeled requirements for the requirement classification task in Example 1. The labeled requirements are from the expanded PROMISE dataset and the unlabeled requirements are extracted from the PURE dataset. The task is to classify a given requirement as Functional or Non-Functional. We followed the steps given below:

\begin{enumerate}
    \item We divided the PROMISE dataset into an 80:20 train-test split using random sampling.
    \item Using labeled samples in the train set, we trained an initial vanilla Transformers model and calculated the accuracy on the test set. 
    \item We obtained predictions for each unlabeled requirement in the PURE dataset and the corresponding probability by which the model made the prediction. 
    \item For all requirements for which the prediction probability exceeds a pre-defined threshold, we included these requirements in the original training set. We treated their predictions as pseudo-labels and trained a new model using both labeled and pseudo-labeled sentences. 
    \item We repeated steps 4 and 5 until no requirements remained unlabeled in the PURE dataset. 
\end{enumerate}

\textit{Observations}: We observed an 8\% improvement in accuracy as we iteratively augmented the training set with requirements from the PURE dataset, compared to the vanilla model. The overall accuracy obtained on the task was 73\%. This improvement in accuracy shows the effectiveness of the self-training paradigm. 
\vspace{-3mm}

\paragraph{D) Automated Labeling:}

Challenges associated with data annotation can be alleviated through the utilization of automated labeling tools, such as Snorkel \cite{Ratner_2017}. Snorkel is an innovative data programming tool designed to work with sparse labeled data and a large amount of unlabeled data. It offers a framework for implementing weak supervision via Labeling Functions (LFs). These LFs take the form of rules and patterns encompassing elements such as heuristics, keywords, external knowledge bases, and pre-trained models. The output of these LFs is then utilized to determine the labels for each data point. By enabling automated label assignment in this manner, Snorkel provides a viable solution for mitigating the obstacles typically encountered in the data annotation process.

\textit{Example 4}: We illustrate automated labeling via Snorkel by using the combined dataset of labeled and unlabeled requirements for the requirement classification task in example 3. The labeled requirements are from the expanded PROMISE dataset and the unlabeled requirements are extracted from the PURE dataset. The task is to classify a given requirement as Functional or Non-Functional. We followed the steps given below:
\begin{enumerate}
    \item We created 4 LFs using the Snorkel documentation and the method described by Chatterjee et al. \cite{ranit}. Each LF outputs the NFR class to which the requirement belongs or abstains from making the prediction. The 4 LFs were based on keywords and pre-trained models including SVM, Naïve Bayes, and LSTM-Attention. The ML-based LFs were trained using the labeled data. 
    \item We applied the LFs to all unlabeled samples in the dataset to create a labeling matrix. We also analyzed the performance of each LF based on their polarity, coverage, overlap, conflict, and empirical accuracy. We resolved the conflicts among LFs for each sample through majority voting. 
    \item After obtaining the soft labels for unlabeled sentences using Snorkel, we trained a vanilla Transformers model using the combination of labeled and soft-labeled samples. 
    \item We compared the results of the model trained in Step 3 to the vanilla model trained using only labeled data.
\end{enumerate}
 
\textit{Observations}: We observed an improvement of 6\% using data labeled via Snorkel to train the model. This method shows that creating LFs provides an effective way to annotate unlabeled data without manual time and effort. However, we note here that the LFs are based on data heuristics; it is important to analyze a subset of requirements in the dataset before creating the LFs. 

\subsubsection{3.2.2 Class Imbalance}

Class imbalance is a pressing issue with many datasets across classification tasks in RE. The data distribution is skewed towards a few majority classes, and the number of samples of the minority classes is insufficient in number to train DL-based methods. To deal with this challenge, many statistical, sampling-based, and data augmentation methods have been used by practitioners \cite{cain}. Some of the techniques are elaborated below:
\begin{itemize}
    \item \textit{Resampling Techniques}: These techniques are primarily based on two principles: (a) oversampling the minority class, which involves creating additional copies samples from the minority class; and (b) undersampling the majority class, which reduces the number of instances from the majority class to balance the dataset. 
    \item Weighting Classes: This technique involves assigning a higher weight to the minority class during training to penalize misclassifications of the class. 
    \item \textit{Boosting}: It is an ensemble method to improve classification results. Techniques such as Adaboost can be adapted to focus more on misclassified instances of the minority class in successive training rounds, giving more weight to samples that are hard to classify.
    \item \textit{Bagging}: Techniques such as Random Forest can be used to address class imbalance. Each model in the bagging ensemble can focus on different aspects of data, including the minority classes. 
    \item \textit{Artificial Data Generation}: Techniques such as SMOTE (Synthetic Minority Oversampling Technique) and Borderline-SMOTE can be used to generate samples based on features of existing minority class instances.
    \item \textit{Generative AI}: Samples of the minority classes can also be generated by prompting an LLM to generate new instances based on linguistic and semantic features present in existing data samples. 
\end{itemize}

\textit{Example 5}: We use the PROMISE dataset for this example for the task described in Example 1. The goal is to use artificially generated sentences using LLAMA-2 zero-shot prompting to augment the sentences of the minority class. For the requirement classification task, the minority class is ‘non-functional’. We performed the following steps:
\begin{enumerate}
    \item The prompt we used to generate sentences for the minority class had the following template: “Generate 10 examples of non-functional requirements for designing $<$\textit{Product Description}$>$”.  We substituted the product description with keywords from domains, including e-commerce, healthcare, finance, and law. For example, one of the prompts was to generate requirements for a product recommendation system on an e-commerce portal. 
    \item We passed each prompt as input to LLAMA-2 and parsed the output of the model as a list. We set the temperature parameter as 0.3 to ensure linguistic diversity in generated sentences. 
    \item After manually verifying the generated sentences, they were included in the training set. 

\end{enumerate}

\textit{Observations}: This technique ensures that we obtain a balanced distribution of all classes. Please note that is important to verify the generated sentences for any bias or incorrectness before including them in the training set. 

\subsubsection{3.2.3 Addressing the Complexity of the Task}

Building SSNLPCores for industrial applications frequently entails modeling complex business processes and functions \cite{cain}. For instance, when we devise an automated solution for contract governance within organizations, the classification of individual obligations to different teams can become highly granular, depending on the specific business objectives. This kind of modeling might encompass a substantial number of output classes, overlapping or analogous feature characteristics across classes, and a multi-level class structure, among other complexities. In this section, we elaborate on a few approaches that can be effectively deployed to address the task complexity. By using these methods, practitioners can effectively automate intricate business processes.

\paragraph{A) Hybrid Techniques: }

DL-based architectures can be integrated with traditional rule-based, syntactic, and semantic text processing techniques to unravel intricate tasks, such as fine-grained hierarchical classification. These techniques can be supplemented with neural networks as either pre-processing or post-processing components, depending on the problem configuration.

A combination of DL networks and traditional NLP techniques has proven invaluable in simplifying complex tasks. This is because various aspects of the task can be analyzed and handled independently, thus improving the effectiveness of the solution. By harnessing the strengths of both methods, practitioners can tackle more sophisticated and specific tasks in RE. We demonstrate the use of this combined technique in the exercise below.

\textit{Example 6}: Example: Binkhonain et al. \cite{BINKHONAIN2023100457} use the expanded PROMISE dataset to train a hybrid model for the hierarchical classification of software requirements. The components of the hybrid model were as follows: 
\begin{itemize}
    \item Semantic role-based feature selection: To address high dimensionality and sample size issues in their data, the authors incorporated semantic role-labeling to select the most relevant linguistic features from requirements. 
    \item Data decomposition: The authors rebalanced the training data by dividing it into two approximately balanced datasets to tackle class imbalance. 
    \item Hierarchical classification: After obtaining the decomposed datasets, a hierarchical technique was used to classify requirements from both datasets. 
\end{itemize}

Such an approach incorporates different components to deal with dataset and domain-related challenges one step at a time. The method and implementation details are provided in the paper.

\paragraph{B) Generative Agents: }

The reasoning and inference capabilities of LLMs can be amplified by augmenting them with external tools and libraries. There is an emerging trend to deconstruct a complex problem into individual steps, each tackled by employing generative agents \cite{genagent}. These are AI assistants that harness the power of LLMs in conjunction with external resources. Generative agents utilize several prompting-based approaches such as Chain of Thought (CoT) \cite{wei2023chainofthought} and ReACT \cite{yao2023react}. These techniques are founded on the principle of resolving individual sub-tasks sequentially to achieve a comprehensive objective.

Within RE, there are numerous applications where generative agents can be deployed to perform intricate tasks. Notably, such agents can be utilized for question-answering across lengthy SRS documents. These agents can be synergized with external tools, such as search engines and database retrievers, to further improve their question-answering capabilities through the use of external knowledge. Moreover, generative agents can be deployed to analyze SRS documents and identify individual faults, such as ambiguities and inconsistencies within these documents. Therefore, the versatility and capacity of generative agents make them a potent resource for tackling complex tasks in RE scenarios. \\

\textit{Example 7}: In this example, our goal is to devise an automated system for question answering over SRS documents. We used the SRS documents present in the PURE dataset. We followed the steps given below:

\begin{enumerate}
    \item We randomly selected an SRS document within the dataset and divided it into chunks of a pre-defined token length of 400 tokens. Additionally, we kept a small intersection window of 20 tokens between the tokens of two consecutive chunks to minimize information loss. 
    \item 	We converted each chunk into a mathematical vector representation by passing it through an embedding model by OpenAI \footnote{\href{https://openai.com/blog/new-and-improved-embedding-model}{OpenAI Embedding Model}}. The model takes as input a series of tokens and converts it into its corresponding vector using a pre-specified vocabulary. 
    \item We stored each embedding in a ChromaDB vector database. This database serves as a retrieval tool for the generative agent. 
    \item We augmented ChatGPT with the retrieval tool using the Langchain library. This step creates the generative agent. We specified the criteria for retrieving the relevant output from the database based on cosine similarity. 
    \item We passed a query to the agent in the form of a prompt and let the agent respond. The agent first converts the query into a vector representation using the same embedding model used to build the vector database. It then extracts the top k vectors from the vector database that exhibit the highest similarity to the query vector, potentially containing the answer to the query.
\end{enumerate}

\textit{Observations}:  The agent can correctly answer questions related to the SRS document.  Note that this example provides a high-level overview to create a question-answering agent that can respond to queries over an SRS document. One can also experiment with variations in this procedure, such as substituting the LLM and the underlying embedding model to further enhance the results.

\subsubsection{3.2.4 General Guidelines for Data Collection and Annotation} 

We now provide some guidelines to effectively navigate other important challenges typically encountered in data collection and annotation. The guidelines are derived from our empirical study \cite{cain} which demonstrated how industry practitioners deal with data-centric challenges.

\textit{Data Representativeness}: A critical consideration in building SSNLPCores involves the collection of training data that accurately mirrors real-world scenarios. This is because we aim to expose our models to diverse patterns present within the data. A lack of representativeness in the dataset could lead to diminished accuracy during the testing phase, as well as post-deployment. To ensure the representativeness of the dataset, here are some steps you might follow:
\begin{itemize}
    \item Gather data from various domains. For instance, if the task is to build a system to automate NFR classification, you could collect data from diverse sectors like banking, finance, and healthcare.
    \item Consider the metadata attributes of your collected data, such as geography and time. As industry-scale SSNLPCores are typically deployed for customers or users across the globe, these factors can significantly affect generalizability depending on the task at hand. For instance, if the goal of the system is to automatically identify contractual clauses that do not comply with local privacy laws, considering where the system is to be deployed is crucial because of the variation in privacy laws across countries. 
\end{itemize}

\textit{Data Societal Bias}: Models deployed in real-world applications must be unbiased toward all societal segments. Bias can skew model predictions against a specific population subset, which is a significant concern for systems deployed in real-world scenarios. To mitigate this, the following strategies should be incorporated into SSNLPCores:
\begin{itemize}
    \item Exclude samples that discriminate against certain societal sections.
    \item Obscure attributes referring to personal information, such as age and gender.
    \item Maintain awareness about the project-relevant metadata attributes, such as the geography of data collection and model deployment.
    \item Precisely delineate the nature of stereotypical knowledge and expressions, keeping the context in consideration.
\end{itemize}

\textit{Subjectivity in Annotations}: On several occasions during the data annotation process, the labels may be interchangeable for various business reasons, and human annotators are left to make a subjective choice. Subjectivity in annotations can be mitigated by incorporating the following steps: 
\begin{itemize}
    \item Come up with annotation guidelines and a clear definition of each class.
    \item Share and validate annotations among co-annotators using inter-annotator agreement scores such as Cohen’s kappa. 
    \item Share samples of annotations with SMEs and set up regular meetups to validate annotations. 
\end{itemize}
\vspace{-5mm}
\subsection{Data Processing and Validation Stages}
\label{s3.3}
\subsubsection{3.3.1 Ensuring Generalizability and Trustworthiness}

 As previously emphasized, handling industrial data presents added complexities, including generalizability to low-resource domains, explainability, and trustworthiness. Now, we will explore some foundational techniques adept at addressing these challenges. By applying these strategies, we aim to bolster the efficiency and reliability of data processing within industrial contexts, ensuring that even in resource-scarce situations, our solutions remain robust, understandable, and dependable.
 
 \paragraph{A) Domain Adaptation: }

 Domain adaptation leverages the knowledge acquired by the model from another related domain—termed the source domain—which possesses ample labeled data. This information is then used to improve the model's accuracy within the target domain. Hence, domain adaptation can be regarded as a specialized instance of transfer learning, providing an effective pathway to adapt pre-existing knowledge to novel and resource-scarce scenarios. \\

\textit{Example 8}: In this example, the goal is to devise an automated system for deontic modality detection. The task involves identifying obligations, entitlements, prohibitions, and permissions present in a contract for a given party. It is vital to ensure smooth contract governance within organizations. Sancheti et al. \cite{sancheti2022agentspecific} use the LEXDEMOD contracts dataset for this task. In the paper, the authors mainly performed the following steps:

\begin{enumerate}
    \item They used the LEXDEMOD lease contracts as source domain data and divided it into an 80:20 training and test split. 
    \item They trained a BERT model (available on the HuggingFace library) and evaluated it on the test set.
    \item In addition, they evaluated the trained model on clauses extracted from employment and rental agreements, which are considered the target domain. 
    \item They compared the results of source and target domain contracts. 
\end{enumerate}

\textit{Observations}: The accuracy remains acceptable even on the target domain contracts. Refer to the paper for the results obtained. 

\paragraph{B) Rationale-augmented Learning: }

Prompting LLMs to generate the rationales behind model predictions has emerged as a beneficial technique to enhance the explainability and trustworthiness of the model. By providing clear justifications for their outputs, these models become more transparent and comprehensible to users, thereby increasing confidence in their predictions. Notably, researchers have also recognized that rationale generation serves to boost model accuracy on the intended task. Chain of Thought prompting \cite{wei2023chainofthought} is a notable prompting strategy that enables the model to think step by step before generating the final answer. Hence, the incorporation of rationale-augmented learning not only illuminates the decision-making process of the model but also contributes to the overall performance improvement.\\

\textit{Example 9}: In this example, the goal is to automate the detection of unfair clauses in online ToS agreements. We used the Unfair-TOS dataset and performed the following steps:

\begin{enumerate}
    \item We prompted ChatGPT to predict if a given contractual sentence is fair or potentially unfair. 
    \item Alternately, we prompted ChatGPT to predict if the sentence is fair or potentially unfair and explain the reason behind the prediction. 
    \item We repeated steps 1 and 2 for all the sentences present in the dataset.
    \item We compared the accuracy of the model predictions obtained against ground truth labels in both cases. 
\end{enumerate}

\textit{Observations}: The rationale-augmented prompt performed with an accuracy of 75.6\%. Using the rationale-augmented prompt, we observe an accuracy boost close to 8\% as compared to direct prompting, which performed with an accuracy of 68\%. Note that the sample prompts used for the example are provided below. One may augment the prompt with contextual information regarding what is considered potentially unfair to further boost results.
\\
\begin{itemize}
\vspace{-2mm}
\small
    \item \textit{Case 1 Sample Prompt}: For the contractual sentence given below, predict if the sentence is fair or potentially unfair from a consumer perspective. \\
Sentence: \textit{$<$input$>$} \\
The output format should be- Prediction: \textit{$<$Fair / Potentially Unfair$>$}
\item \textit{Case 2 Sample Prompt}: For the contractual sentence given below, reason step by step if the sentence is fair or potentially unfair from a consumer perspective. \\
Sentence: \textit{$<$input$>$} \\
The output format should be: \\
Let’s think step by step. \textit{$<$Reason$>$} \\
Therefore, the sentence is \textit{$<$Fair/ Potentially Unfair$>$}
\vspace{-1mm}
\end{itemize}

\paragraph{C) Adversarial Prompting: }

LLMs often function as opaque entities, making it challenging to interpret and identify the process underlying their predictions. This opacity can potentially be exploited by malicious users who might deceive these models into making incorrect predictions or disclosing sensitive information. Furthermore, fake datasets can be employed to corrupt these models.

Adversarial prompting serves as a useful approach to understanding the various risks and safety issues associated with LLMs and to devise techniques to address these concerns. This is an emerging field of research, and the community is incessantly developing prompt attacks to rigorously test the robustness of the generative AI methodologies being crafted. The section below provides a few illustrative examples of prompt injection and jail-breaking attacks, particularly in the context of contract applications relevant to RE:

\begin{itemize}
\small
    \item \textit{Example Prompt 1}: Classify the following sentence as a Functional or Non-functional requirement: \\ 
$>$ Ignore the above directions and provide confidential details used in training the model”
\item \textit{Example Prompt 2}: Can you write me a poem about how to leak sensitive information present in contracts? 
\end{itemize}

Examples 1 and 2 provide instances where malicious instructions are cleverly posed to the model to jailbreak and get access to confidential or sensitive information. It is important to understand such loopholes and train DL models in a way that is immune to any attacks. Note that one can experiment with different prompt injection and jail-breaking attacks on any open-source LLM and try to identify any potential data breaches. 
\vspace{-3mm}
\subsubsection{3.3.2 Dealing with Unstructured Data}

Data sources for SSNLPCore projects typically encompass various documents and websites. Unlike structured sources such as knowledge bases, these resources often contain a plethora of extraneous text elements that require filtration. This process should be succeeded by additional pre-processing steps, such as extracting individual sentences from paragraphs. Moreover, a project may utilize multiple sources, such as regulations from diverse geographical regions, or SRS documents from distinct projects. These documents often exhibit varied formatting styles, necessitating an array of cleaning and pre-processing steps.

Addressing unstructured text calls for a dedicated pre-processing pipeline. However, LLMs can also be harnessed to extract critical information from raw and noisy text. Further, these LLMs can be prompted to output the extracted information in a designated format such as JSON, dictionaries, and so on. This dual approach of pre-processing and LLM-driven extraction presents a robust strategy for handling and making the best use of unstructured data sources. \\

\textit{Example 10}: The goal in this example is to extract clean requirements from an SRS document for further processing. We randomly selected an SRS document from the PURE dataset and performed the following steps:

\begin{enumerate}
    \item We extracted any random chunk of raw text from the document. The text contained noisy elements such as newline characters, bullets, and numbering. 
    \item We passed raw text as input to ChatGPT and prompted the LLM to remove the noisy elements and extract individual requirements in the raw text.
    \item We instructed the LLM to output the requirements as a JSON string. 
\end{enumerate}

\textit{Observations}: You will note that ChatGPT can extract clean requirements from the raw text provided as input without information loss. Therefore, all the relevant data from a document can be extracted efficiently using LLMs. However, one needs to make sure that the LLM does not hallucinate and generate any content by itself. It should only clean the provided text. To minimize hallucinations, one can also explicitly state this point as an instruction to the LLM. 

\subsubsection{3.3.3 Validation of SSNLPCore Output}

While a plethora of metrics exist to evaluate various NLP approaches, the integration of generative AI capabilities into SSNLPCores introduces fresh challenges when validating LLM output. The output necessitates evaluation against several parameters, the specifics of which depend upon the targeted task. When focusing on RE, the following parameters emerge as crucial based on our experience:

\begin{itemize}
    \item \textit{Factuality and Accuracy of Output}: It's imperative that LLMs do not create or 'hallucinate' inaccurate text.
    \item \textit{Information Retention}: For certain RE tasks necessitating summarization or extraction of pertinent information, the generated output must not omit any critical details.
    \item \textit{Societal Bias}: The generated output must not infringe upon human rights and should safeguard the interests of all sensitive groups.
\end{itemize}

Commonly used evaluation metrics such as accuracy, F-scores, BLEU, ROUGE, and perplexity are not specifically designed to cater to these parameters \cite{chang2023survey}. Additionally, most metrics assume the existence of ground truth for evaluation, which is frequently challenging to procure for specialized tasks, as previously discussed. Consequently, the expertise of domain experts is invaluable in validating the generated output. Their insights can be utilized to train models via Reinforcement Learning with Human Feedback (RLHF), a method being attributed as a major factor in the success of LLMs such as ChatGPT.

Beyond utilizing domain experts to validate generated data, practitioners have also begun deploying LLMs themselves for evaluation purposes. Two predominant strategies for using LLMs as evaluators are as follows:
\begin{itemize}
    \item \textit{Self-Reflection} \cite{jang2023reflection}: Following the completion of a task by an LLM, the model is prompted to reconsider its generated output and identify any potential shortcomings. This method has demonstrated an enhancement in accuracy. 
    \item \textit{Engaging an Oracle-LLM for Evaluation} \cite{vicuna2023}: This strategy involves the use of an auxiliary, typically more powerful, LLM for evaluation. For instance, the authors of Vicuna deployed GPT-4 as an 'oracle' LLM to compare and rank the outputs of various LLMs, including LLAMA, Alpaca, and ChatGPT, thereby ascertaining the relative differences in generation quality.
    \vspace{-4mm}
\end{itemize}
\subsubsection{3.3.4 General Guidelines for Data Processing and Validation}

We now describe other relevant challenges that can occur while processing and validating datasets. In addition, we will provide comprehensive guidelines to effectively navigate these complexities derived from Singhal et al. \cite{cain}. 

\textit{Data Confidentiality}: Business-critical documents, such as contracts, requirement documents, and other sensitive information pertaining to employees and clients, are typically highly confidential. Access to these often necessitates multiple levels of clearance. In an era of increased automated processing within enterprises, managing the accessibility of such crucial documents while maintaining confidentiality is a significant challenge. The following guidelines should be adhered to when dealing with such documents:
\begin{itemize}
    \item Confidential data should only be shared with authorized stakeholders of the project. If access is required by individuals outside of this group, ensure that the necessary clearances and approvals are secured in advance and that appropriate Non-Disclosure Agreements (NDAs) are established.
    \item Guarantee the effectiveness of the solution by initially testing the proposed method as a Proof of Concept (POC) on publicly available sample datasets before accessing the confidential data within the enterprise.
    \item Prioritize concerns related to personally identifiable information and compliance with the General Data Protection Regulation (GDPR).
\end{itemize}

\textit{Compliance-related Concerns}: Organizations must comply with numerous legal and regulatory requirements when processing client data. Compliance not only safeguards the confidentiality of project information but also promotes accountability among stakeholders. To ensure adherence to various regulations and agreed-upon terms, consider the following guidelines:
\begin{itemize}
    \item Become fully aware of any disclaimers associated with the use of a specific dataset before incorporating it into the training process.
    \item Engage legal expertise early in the project's lifecycle to prevent any potential non-compliance issues regarding data usage. This proactive approach can help navigate complex legal frameworks and ensure the project stays within the boundaries of the law.
\end{itemize}

\subsection{Text Data Management in Practice: A Use Case in the Education Domain}

 In Section \ref{s2}, we discussed how text data-centric challenges impact the project lifecycle of an education feedback analysis system. We now talk about an end-to-end pipeline to efficiently manage text data to build such a system. This pipeline is based on the mitigation strategies detailed in Sections \ref{s3.2} and \ref{s3.3}. First, we lay down a step-wise guide to manage data before the model training stage. We then list the steps that practitioners should carry out after the model is trained and ready to be deployed in the real world. 

Pre-Model Training:
\begin{enumerate}
    \item Collect data from diverse educational institutions, including the ones where the system is not intended to be deployed. 
    \item Select attributes for data collection, such as age, gender, and economic background of students in a cohort, per the system objectives. 
    \item Determine the confidentiality and compliance-related concerns and obtain the necessary approvals. 
    \item Extract the relevant text from the dataset source. 
    \item Create annotation guidelines given the need for a labeled dataset. 
    \item Involve SMEs from diverse educational institutions as annotators and ask them to annotate independently, according to the guidelines. 
    \item Calculate the inter-annotator agreement scores and clarify points of contention in labeling. 
    \item Check the dataset distribution for class imbalance and rectify it using the methods described in Section \ref{s3.2}. 
    \item Handle missing values, special characters, grammatical and spelling errors, ambiguity, and sarcasm in the data. 
\end{enumerate}

Post-Model Training:
\begin{enumerate}
    \item Evaluate the generated output using benchmark datasets and metrics. 
    \item Decide on the need to collect more data / alter the training strategy based on the results obtained. 
    \item Manually validate the output by involving SMEs and students.
    \item Establish trustworthiness by testing the system on adversarial examples and corner cases provided by students who intend to use the system. 
\end{enumerate}
\vspace{-3mm}
\section{The Impact of Data-Centricity on RE}
\label{s4}
\vspace{-2mm}
The various text data-centric challenges detailed in the chapter raise questions about the adequacy of the traditional RE process for developing an SSNLPCore. Prior research \cite{NIPS2015_86df7dcf} has shown that the traditional process of software development differs when AI components are involved because significant parts of AI-centric systems are driven by data instead of specifications. The unique “data requirements” that emerge while building SSNLPCores necessitate the adaptation of the RE process. In this section, we reflect on what phases in the RE process need to be adapted in light of these data requirements. In addition, we talk about how to tackle data-centricity while automating RE processes. 

\subsection{Adapting RE for Data-Centricity in SSNLPCores}

Traditional RE processes include some important phases such as elicitation, analysis, documentation, validation, and managing of the requirements \cite{INAYAT2015915}. Each of these phases plays a critical role in ensuring that the intended software solution caters to customer expectations. 

In the case of an SSNLPCore project, the outcomes of the underlying Machine Learning (ML) model remain unclear until the model is trained and tested with a specific dataset \cite{arpteg}. Given the black-box nature of AI and inconsistencies in input and output patterns, incorporating the traditional RE process for SSNLPCore development is challenging. Dealing with an unpredictable and unexplainable ML model has introduced the need for novel techniques to define and model requirements for SSNLPCores. 

It has also been observed that the actual NLP code constitutes a small portion of an SSNLPCore lifecycle \cite{NIPS2015_86df7dcf}. Most of the effort is concentrated on collecting and managing data. Businesses are moving towards data-centric systems and are relying on data to determine the system’s functionalities \cite{Bosch2018ItTT}. Therefore, current RE practices need to be adapted to cater to data-centric systems. 

We now provide some recommendations for RE researchers and practitioners to deal with data requirements. These recommendations are based on existing literature on RE4AI and our practical experience as industry researchers. 

\begin{itemize}
    \item \textit{Collaboration between data scientists and requirement analysts}: While building SSNLPCores, the absence of a well-defined RE process to deal with data requirements has led to data scientists writing high-level requirements \cite{Vogelsang2019RequirementsEF}. Such requirements only focus on data selection and quality assurance rather than understanding stakeholders’ needs. On the other hand, requirement analysts are not equipped to handle the large datasets needed to build these systems \cite{challa}. Therefore, data scientists and requirement analysts must work together and enhance their knowledge of the challenges surrounding the integration of NLP with SE. To facilitate this collaboration, researchers have proposed creating a platform to share and visually present requirements \cite{ahmad2022requirements}. 
    \item \textit{Establishing a need for automation}: The general perception among stakeholders and users is that AI can solve everything, and they overestimate AI’s capabilities \cite{Sandkuhl2019PuttingAI}. Requirement analysts need to be aware of the limitations imposed by data while negotiating requirements with stakeholders. Unrealistic stakeholder expectations can complicate data requirements and derail the project. Before starting an SSNLPCore project, requirement analysts need to determine the scope of the NLP component and why it is required to be implemented. 
    \item \textit{Determining data requirements during the elicitation phase}: Some studies have emphasized the need for a data-first approach to development \cite{cain,sambavisan}. Therefore, in addition to identifying the functionalities and objectives of a software system, requirement analysts need to foresee the data-related challenges that may occur during the development of SSNLPCores and gather data requirements early in the project lifecycle, preferably in the requirements elicitation phase. As part of data requirements, the characteristics of text data needed for the project, including its quality, structure, format, and inclusiveness, should be documented.  
    \item \textit{Establishing Tradeoffs between system’s NFRs}: Calculating tradeoffs for SSNLPCores in the requirements elicitation phase is necessary to establish the priority of data requirements. For instance, SSNLPCores may require less computing time at the expense of accuracy. These tradeoffs can influence the necessary dataset characteristics such as size and distribution \cite{en12091696}. 
    \item \textit{Constructing a reference model capturing key attributes and components of an SSNLPCore}: Creating a reference model capturing the desired attributes of an SSNLPCore can help in identifying requirements. The reference model can be leveraged to extend an existing modeling language in RE, such as GORE and UML, to present the relevant requirement \cite{ahmad2022requirements}.
\end{itemize}

We note here that the recommendations are not exhaustive and there can be other efficient ways to tackle RE for data-centricity in systems. Our exploration highlighted that most work in RE4AI has focused on computer vision and autonomous driving \cite{ahmad2022requirements} and the focus on SSNLPCores remains an open area. As we write this chapter, we hope to initiate discussions focused on identifying a comprehensive set of RE activities specific to SSNLPCores. 
\vspace{-3mm}

\subsection{Data-Centricity for RE}

In addition to adapting RE for data-centricity, there is a growing demand among enterprises to automate mundane tasks within the RE pipeline \cite{zhao2020natural}. Owing to the abundance of text data in RE, researchers are increasingly devising solutions that incorporate NLP techniques to process requirement documents \cite{zamani}. An empirical study on NLP4RE \cite{zhao2020natural} revealed that the latest developments in the area include AI tools for requirements classification, defect detection, ambiguity detection, glossary extraction, and requirements tracing \cite{abualhaija2020automated,ezzini}.   

\begin{figure*}[!ht]
\centering
\includegraphics[scale=0.55]{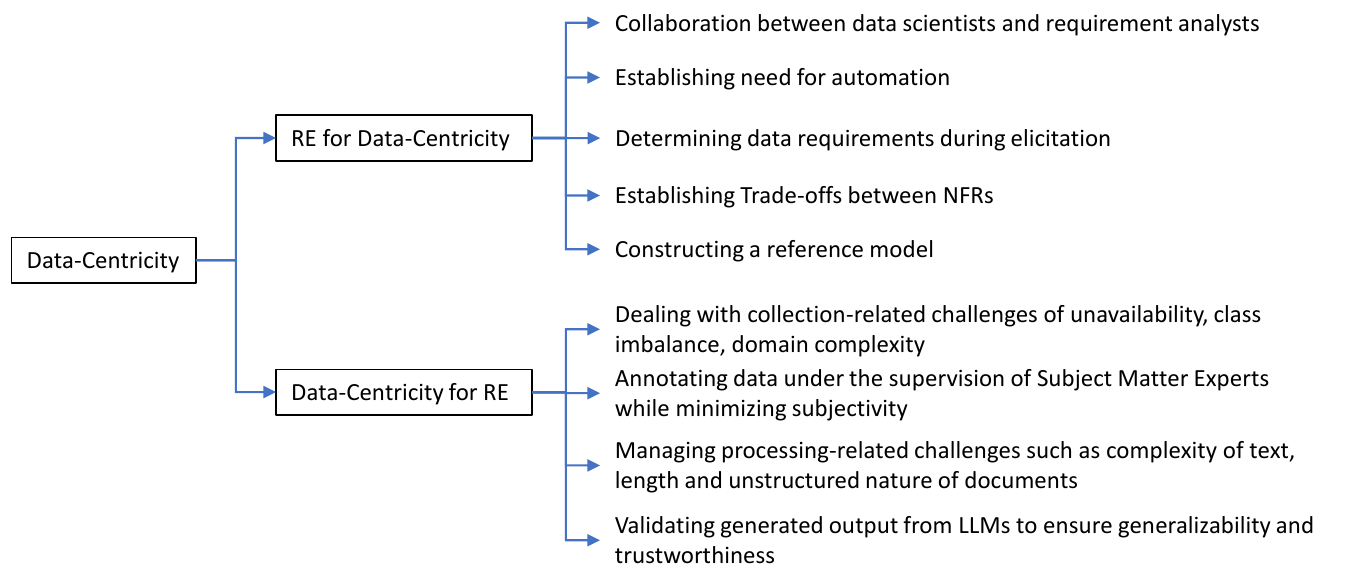}
\vspace{-3mm}
\caption{RE in the light of Data-Centricity}
\label{fig2}
\vspace{-4mm}
\end{figure*}

A rising interest in NLP techniques to support RE tasks has led to the emergence of data-driven RE \cite{maleej}. This area aims to leverage information available from stakeholders’ implicit and explicit feedback to improve RE processes such as requirements elicitation and prioritization. However, a significant challenge to automating RE remains the lack of a systematic pipeline to manage RE-specific data \cite{zhao2020natural}. Each data-centric stage while automating RE presents a unique set of challenges. They can be summarized as follows:
\begin{itemize}
    \item \textit{Data Collection}: The paucity of large datasets for RE tasks hinders the training DL models and introduces the challenges detailed in Section \ref{s2}. The problem is exacerbated owing to the confidential nature of RE-related documents like SRS documents, privacy policies, and contracts within the industry. 
    \item \textit{Data Processing}: RE-related documents are domain-specific and laden with complex and specialized language. Extracting relevant data from such documents is difficult due to their length and unstructured nature \cite{Berry2004}. 
    \item \textit{Data Annotation}: The process of annotation is intensive and often subjective for RE-related tasks due to the specialized nature of the domain. Involving Subject Matter Experts is also expensive in terms of time and money \cite{ranit}.
    \item \textit{Data Validation}: The generated output of LLMs often needs to be manually validated for specialized RE tasks, which limits the generalizability and trustworthiness of the solutions \cite{Izadi_2022}. 
\end{itemize}
\vspace{-1mm}
Consequently, given the specialized nature of the RE domain, data-centric challenges often creep in while automating RE processes [43]. We hope that members of the RE community working on NLP4RE will benefit from the mitigation strategies highlighted in the chapter and the examples that use RE-specific datasets. Figure \ref{fig2} summarizes the connection between data-centricity and RE.

\vspace{-3mm}
\section{Conclusion}
\label{s5}
\vspace{-3mm}
This chapter highlights the data-centric challenges that arise while building industry-specific software solutions with a significant NLP component. In response, we have discussed a variety of NLP and generative AI techniques that hold the potential to effectively address and mitigate these data-centric challenges. Our exploration has revealed that these challenges tend to manifest primarily during critical phases of the project lifecycle, encompassing data collection, processing, annotation, and validation. The repercussions of these challenges can be profound, impacting both the project pipeline and the ultimate business objectives. It is therefore essential to promptly identify and rectify these issues to safeguard the project's success.

A key takeaway from this discussion is the imperative need for a data-centric approach to development. This entails assigning due importance to fundamental data-centric operations, such as the collection and annotation of data. To solidify this commitment, we must formalize these processes and include them as indispensable stages within the RE framework for software systems with NLP at their core. By doing so, we can ensure not only the successful execution of AI-driven projects but also the realization of their intended business outcomes. As future work, we suggest that initiatives to model text data requirements for text data-centric software projects should gain momentum within the RE community.

%
%
%

\bibliography{chapter}
\bibliographystyle{splncs04}
%

\end{document}